\documentclass[aps,floatfix,prl,twocolumn,groupedaddress,showpacs,showkeys,tightenlines,superscriptaddress,nofootinbib]{revtex4-2}

\usepackage{color}
\usepackage{lipsum}
\usepackage{amssymb,amsmath,MnSymbol}
\usepackage{graphicx,slashed}
\usepackage{subfigure}
\usepackage[bottom]{footmisc}
\usepackage{dsfont}
\usepackage{multirow}
\usepackage{hyperref}

\newcommand{\bs}{b_{\rm S}}
\newcommand{\eq}{\begin{eqnarray}}
	\newcommand{\en}{\end{eqnarray}}

\newcommand{\be}{\begin{equation}}
\newcommand{\ee}{\end{equation}}
\newcommand{\ba}{\begin{eqnarray}}
\newcommand{\ea}{\end{eqnarray}}
\newcommand{\la}{\label}
\newcommand{\mqav}{m_{\rm q}^{\rm av}}
\newcommand{\mql}{m_{{\rm q},l}}
\newcommand{\mqs}{m_{{\rm q},s}}

\newcommand{\zs}{Z_{\rm S}}
\newcommand{\zm}{Z_{\rm m}}

\newcommand{\rs}{r_{\rm S}}

\newcommand{\gs}{g_{\rm S}}
\newcommand{\txts}{\textstyle}
\newcommand{\fs}{f_{\rm S}}

\newcommand{\dbars}{\overline{d}_{\rm S}}
\newcommand{\dbarm}{\overline{d}_{\rm m}}

\newcommand{\bbars}{\overline{b}_{\rm S}}
\newcommand{\bbarm}{\overline{b}_{\rm m}}
\newcommand{\dm}{d_{\rm m}}

\newcommand{\m}{\mathbf}

\newcommand{\ds}{d_{\rm S}}
\newcommand{\bg}{b_{\rm g}}
\newcommand{\rrm}{r_{\rm m}}

\begin{document}
	
	\title{Nucleon Sigma Terms with $N_f = 2 + 1$ O($a$)-improved Wilson fermions}
	
\author{A. Agadjanov}
	 \affiliation{PRISMA$^+$ Cluster of Excellence \& Institut f\"ur Kernphysik,
	 Johannes Gutenberg-Universit\"at Mainz, D-55099 Mainz, Germany}
\author{D. Djukanovic}
        \affiliation{Helmholtz Institute Mainz, Staudingerweg 18, D-55128 Mainz, Germany}
        \affiliation{GSI Helmholtzzentrum f\"ur Schwerionenforschung, D-64291 Darmstadt,
          Germany}
\author{G.~von~Hippel}
	\affiliation{PRISMA$^+$ Cluster of Excellence \& Institut f\"ur Kernphysik,
	Johannes Gutenberg-Universit\"at Mainz, D-55099 Mainz, Germany}
\author{H. B. Meyer}
	\affiliation{PRISMA$^+$ Cluster of Excellence \& Institut f\"ur Kernphysik,
	 Johannes Gutenberg-Universit\"at Mainz, D-55099 Mainz, Germany}
        \affiliation{Helmholtz Institute Mainz, Staudingerweg 18, D-55128 Mainz, Germany}
\author{K. Ottnad}
	\affiliation{PRISMA$^+$ Cluster of Excellence \& Institut f\"ur Kernphysik,
	 Johannes Gutenberg-Universit\"at Mainz, D-55099 Mainz, Germany}
\author{H. Wittig}
        \affiliation{PRISMA$^+$ Cluster of Excellence \& Institut f\"ur Kernphysik,
	 Johannes Gutenberg-Universit\"at Mainz, D-55099 Mainz, Germany}
        \affiliation{Helmholtz Institute Mainz, Staudingerweg 18, D-55128 Mainz, Germany}
	
\date{\today}

\begin{abstract}
We present a lattice-QCD based analysis of the nucleon sigma terms using gauge ensembles
with $N_f = 2 + 1$ flavors of ${\cal O}(a)$-improved Wilson fermions,
with a complete error budget concerning excited-state contaminations, the
chiral interpolation as well as finite-size and lattice spacing effects.
We compute the sigma terms determined directly from the matrix
elements of the scalar currents. The chiral interpolation is based on SU(3)
baryon chiral perturbation theory using the extended on-mass shell
renormalization scheme. For the pion nucleon sigma term, we
obtain $\sigma_{\pi N} = (43.7\pm3.6)$\;MeV, where the error
includes our estimate of the aforementioned systematics. The tension with extractions based on
dispersion theory persists at the 2.4-$\sigma$ level. For the strange
sigma term, we obtain a non-zero value, $\sigma_s=(28.6\pm9.3)$\;MeV.
\end{abstract}
	
\maketitle
	
\paragraph{\bf Introduction.}
	
The scalar matrix element of the nucleon
is an important observable, and plays a crucial role in
interpreting the results of dark-matter direct-detection
experiments. Especially appealing candidates for cold dark matter are
weakly interacting massive particles (WIMP), as they naturally
reproduce the observed relic abundance of dark matter through
annihilation processes in the early universe. In particular for
Higgs-portal models, in which the WIMP-nucleus interaction is mediated
by the Higgs boson, the spin-independent cross-section for
WIMP-nucleus recoil experiments is sensitive to the values of the
scalar matrix element \cite{Jungman:1995df}. The light-quark scalar
matrix element\footnote{We take the nucleon at rest and use the state
normalization $\langle N\,\vec p'\,s'|N\,\vec p \,s\rangle = (2\pi)^3
\delta^{ss'}\delta(\vec p-\vec p')$. Also, throughout this work we
assume exact isospin symmetry.}
        \be \label{eq:sigpiNdef} 
        \sigma_{\pi N} \equiv m_l \,\langle N|\bar u u + \bar d d| N\rangle =
	m_l\; ({\partial m_N}/{\partial m_l}),
        \ee
where $m_l \equiv (m_u+m_d)/2$, also known as the pion-nucleon sigma
term, is of special interest. Phenomenologically, $\sigma_{\pi N}$ is
accessible via $\pi N$-scattering amplitudes at the Cheng-Dashen point
\cite{Cheng:1970mx}.  Historically, the value for $\sigma_{\pi
  N}\sim 45 \ {\rm MeV}$ derived in \cite{Gasser:1990ce} was prevalent
for a long time, a value compatible with most lattice
determinations. However, new analyses using constraints from pionic
hydrogen and deuterium led to a much larger value of $\sigma_{\pi
  N}=59.1(3.5)\ {\rm MeV}$\, \cite{Hoferichter:2015dsa}, consistent with
  the EFT analysis of \cite{Alarcon:2011zs} and in agreement with
  \cite{RuizdeElvira:2017stg} based on low energy $\pi N$-scattering (see
Ref.~\cite{Hoferichter:2015hva} for a review).  By contrast, lattice
calculations for $\sigma_{\pi N}$ \cite{Durr:2011mp, Bali:2012qs,
  Shanahan:2012wh, Engelhardt:2012gd, Alexandrou:2014sha,
  Durr:2015dna, Yang:2015uis, Bali:2016lvx, Yamanaka:2018uud,
  Alexandrou:2019brg, Borsanyi:2020bpd, RQCD:2022xux}, discussed in
detail in the FLAG report
\cite{FlavourLatticeAveragingGroupFLAG:2021npn}, have largely
confirmed the lower estimate, while being in tension with the latest
dispersive analysis at the level of 3$-$4 standard
deviations.\footnote{See Refs. \protect\cite{Alvarez-Ruso:2013fza,
  Ren:2017fbv, Lutz:2018cqo, Lutz:2023xpi} for further efforts to
extract $\sigma_{\pi N}$ from collections of lattice data for the
light quark mass dependence of $m_N$.}  Very recently, it
was suggested that the discrepancy is alleviated via an explicit
treatment of $N\pi$ and $N\pi\pi$ excited states in the analysis
~\cite{Gupta:2021ahb}. As a related quantity, the strangeness
matrix element \be \sigma_s \equiv m_s \langle N| \bar ss| N\rangle =
m_s \;({\partial m_N}/{\partial m_s}), \ee a pure sea-quark effect,
has often been discussed together with the pion-nucleon sigma term.
Their linear combination
\be
  \sigma_0  \equiv   m_l \,\langle N|\bar u u	+ \bar d d -2 \bar s s| N\rangle
  = \sigma_{\pi N} - (2{m_l}/{m_s})\sigma_s
\ee
is to first order in $(m_l-m_s)$  proportional to the nucleon-hyperon mass splitting. 
The $\sigma_0$ value inferred from this observation,
assuming a negligible strangeness content $\sigma_s$ of the nucleon,
corresponds to a small value for $\sigma_{\pi N}$.
In~\cite{Alarcon:2012nr} however, corrections to $\sigma_0$ were calculated which
bring the associated $\sigma_{\pi N}$ estimate back into agreement 
with its Cheng-Dashen-theorem based determinations
without the need to invoke a large $\sigma_s$ value.

We perform a direct determination of the nucleon sigma terms from a
lattice calculation of the matrix element of the scalar current.  Our
final estimates are based on a simultaneous chiral, continuum and
infinite volume extrapolation of the pion-nucleon and strange sigma
terms.
We average the individual fits with weights based on the Akaike information
criterion (AIC) \cite{akaike1973second,Akaike:IEEE:1100705}  to
provide a full error budget accounting for variations in the treatment
of excited state contaminations, discretization errors, finite-volume effects and 
the quark-mass dependence.

\paragraph{\bf Simulation details.}	\label{sec:setup}

We employ the  $N_f=2+1$ ensembles \cite{Bruno:2014jqa} generated as part
of the Coordinated Lattice Simulations (CLS) initiative with
non-perturbatively $\mathcal{O}(a)$-improved Wilson fermions
\cite{Sheikholeslami:1985ij} and the tree-level improved
L\"uscher-Weisz gauge action \cite{Luscher:1984xn},
correcting for the treatment of the strange quark determinant using \cite{Mohler:2020txx}.  
Table~\ref{tab:ensembles}  gives details of the ensembles
used in this work. In particular, lattice spacings range from $0.050$\;fm to $0.086$\;fm.

The two-point and three-point functions needed to extract the scalar
matrix elements of the nucleon read
\begin{align}
		C_2(t;{\bf{p}})&=\Gamma_{\alpha\beta} \sum\limits_{{\bf{x}}} e^{-i{\bf{px}}}
		\Bigl\langle\Psi_\beta({\bf{x}},t) \overline{\Psi}_\alpha (0) \Bigr\rangle,\\
		C_3(t,t_s;{\bf{q}})&=\Gamma'_{\alpha\beta} \sum\limits_{{\bf{x,y}}} e^{i{\bf{qy}}}
		\Bigl\langle\Psi_\beta({\bf{x}},t_s) S_q({\bf{y}},t) \overline{\Psi}_\alpha (0) \Bigr\rangle ,
\end{align}
where $S_q$ denotes the scalar density,
\eq \label{eq:O_S}
           S_q = \bar q q, \quad q = u,d,s. 
\en 

The interpolating operator for the proton,
	\begin{equation}
		\Psi_\alpha(x) = \epsilon_{abc}
		\left(\tilde{u}^T_a(x)C\gamma_5\tilde{d}_b(x)\right)\tilde{u}_{c,\alpha}(x)\,,
	\end{equation}
is built using Gaussian-smeared quark fields \cite{Gusken:1989ad} 
	\begin{equation}
		\tilde{q} = (1 + \kappa_{\rm G}\Delta)^{N_{\rm G}} q\,, \qquad q=u,d,
	\end{equation}
	and spatially APE-smeared  gauge links in the covariant Laplacian $\Delta$ \cite{APE:1987ehd}. 
	
The pertinent Wick contractions for the three-point function lead to the
connected and disconnected contributions, $C_3=C_3^{\rm conn}+C_3^{\rm disc}$.
For the connected part, we employ
extended propagators via the ``fixed-sink'' method, requiring additional
inversions for each chosen value of $t_s$ \cite{Martinelli:1988rr}. In order to
reduce the cost of the inversions, we apply the truncated solver method with
bias correction \cite{Bali:2009hu,Blum:2012uh,Shintani:2014vja}. For the
connected part, the polarization matrices $\Gamma',\Gamma$ read
	\begin{align}
		\Gamma'=\Gamma = {\txts\frac{1}{2}}(1+\gamma_0) (1+i \gamma_5 {\gamma_3}).\label{eq:polarization}
	\end{align}
	
The disconnected three-point function is constructed from the quark
	loop ${L}^q$ and the nucleon two-point function
	\begin{eqnarray}
		C^{\rm disc}_{3}(t,t_s;{\bf{q}}) &=& 
		\Bigl\langle
		e^{-i\m{q}\m{x}} {L}^q(\m{q},z_0)\cdot
		C_2(\m{p}^\prime,y_0,x;\Gamma') \Bigr\rangle ,
	\end{eqnarray}
	where
	\begin{eqnarray}
L^q(\m{q},z_0)  &=&
-\sum_{\m{z}\in\Lambda} e^{i\m{q}\cdot\m{z}} {\rm Tr}\left[D_q^{-1}(z;z)\
\mathds{1}\right]\,.
\label{eq:strange_loop}
	\end{eqnarray}
	Note that for forward scalar matrix elements ($\m q=0$), the vacuum
	expectation value of the current insertion must be subtracted,
	\begin{align}
		C^{\rm disc}_{3}(t,t_s;{\bf{0}}) &= 
		\Bigl\langle
		{L}^{q}(\m{0},z_0)\cdot
		C_2(\m{p}^\prime,y_0,x;\Gamma') \Bigr\rangle \nonumber\\
		&- \Bigl\langle
		{L}^{q}(\m{0},z_0)\Bigr\rangle \cdot\Bigl\langle
		C_2(\m{p}^\prime,y_0,x;\Gamma') \Bigr\rangle.
	\end{align}
Additionally, we improve the signal by averaging over all three different polarizations
	\eq
	\Gamma'_i={\txts\frac{1}{2}}(1+\gamma_0) (1+i \gamma_5 {\gamma_i}),\quad i = 1,2,3,
	\label{eq:polarization_disc}
	\en
	and by averaging over forward and backward propagating nucleons. Traces
	over the  quark loops are  estimated stochastically using
	four-dimensional noise vectors $\eta$. 
We improve the precision of the
quark loops using  a variation of the
frequency splitting method \cite{Giusti:2019kff} that combines the
one-end-trick \cite{McNeile:2006bz}  with a generalized hopping parameter
expansion \cite{Gulpers:2013uca} and
hierarchical probing \cite{Stathopoulos:2013aci} (for more details see App. C Ref.~\cite{Ce:2022eix}).

\begin{table}[!t]
	\begin{ruledtabular}
		\begin{tabular}{cccccc}
			ID  & $t_0/a^2$ & $M_\pi [{\rm MeV}]$ &
			$M_K [{\rm MeV}]$ &  $M_\pi L$  &
			$ \{t_{\rm s}^{\rm min} ,t_{\rm s}^{\rm max}, 
			t_{\rm s}^{\rm step} \}/a$ \\\hline
			H102 & 2.860  &  352  &  437 & 4.93 & \{4,17,1\}
			\\
			N101 & 2.860  &  278  &  461 & 5.83 & \{4,17,1\}\\
			H105 & 2.860  &  277  &  462 & 3.88 & \{4,17,1\}\\
			C101 & 2.860  &  219  &  470 & 4.59 & \{4,17,1\}\\\hline
			S400 & 3.659  &  349  &  440 & 4.32 & \{4,22,2\}\\
			N451 & 3.659  &  286  &  461 & 5.31 & \{4,20,2\}\\
			D450 & 3.659  &  215  &  475 & 5.33 & \{4,20,1\}\\
			D452 & 3.659  &  154  &  482 & 3.80 & \{4,20,2\}\\\hline
			N203 & 5.164  &  346  &  442 & 5.40 & \{4,24,2\}\\
			S201 & 5.164  &  288  &  467 & 3.00 & \{4,22,2\}\\
			N200 & 5.164  &  284  &  463 & 4.43 & \{4,22,2\}\\
			D200 & 5.164  &  200  &  480 & 4.16 & \{4,22,2\}\\
			E250 & 5.164  &  128  &  489 & 4.00 & \{4,22,2\}\\\hline
			N302 & 8.595  &  344  &  450 & 4.17 & \{4,28,2\}\\
			J303 & 8.595  &  257  &  474 & 4.14 & \{4,28,2\}\\
			E300 & 8.595  &  174  &  490 & 4.22 & \{4,28,2\}\\
		\end{tabular}
	\end{ruledtabular}
		\caption{Details of CLS ensembles used in this work. The pion
		and kaon masses are taken from~\cite{Ce:2022kxy} and the lattice
		spacings
		from~\cite{Bruno:2016plf}. Using Eq. (\ref{eq:t0phys}) the largest source-sink separations
		correspond to 1.4 fm and 1.5 fm for the two finer and coarser
		lattices, respectively.
	}
		\label{tab:ensembles}
	\end{table}
	
\begin{figure*}[!t]
	\centering
	\includegraphics[width=0.9\textwidth]{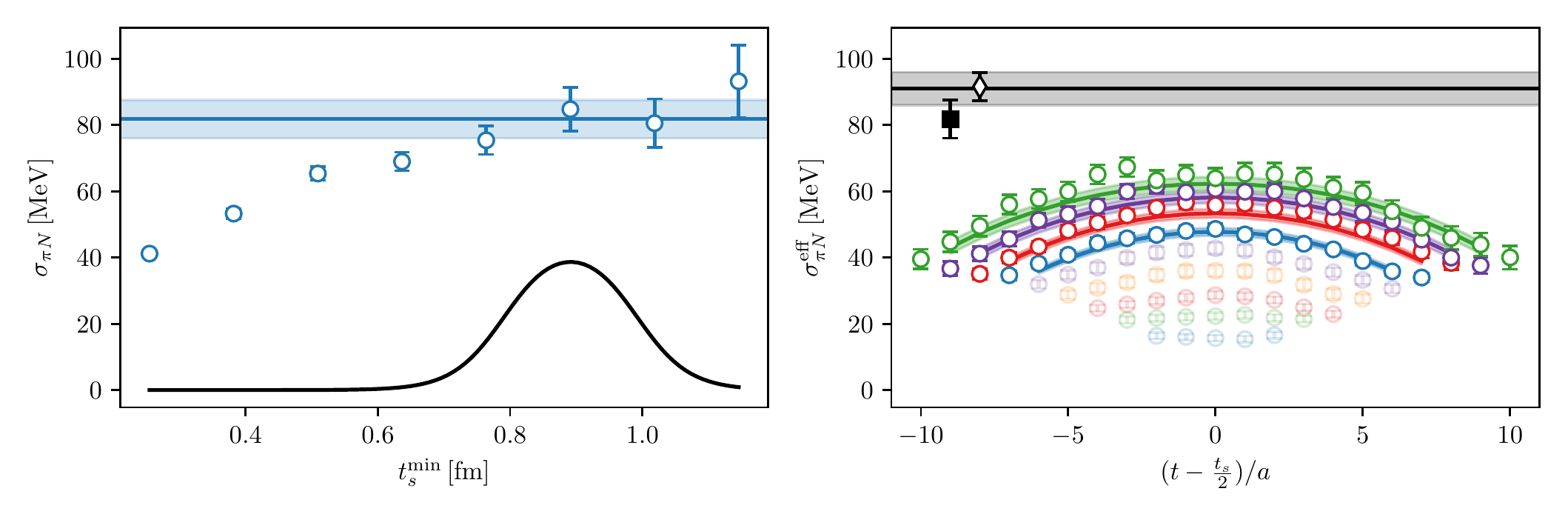}
	\caption{Left: Results of linear fits to the summed
	correlator on ensemble D200 with the starting time slice given on the x-axis. The blue
	shaded area is the weighted average using Eq.
(\ref{eq:weight_function}) shown as a black line in the bottom of the plot, for
the particular choice of parameters from Eq. (\ref{eq:choice_window}). Right:
Fit result of an explicit two-state fit to the effective form
factor. The gray band represents the result for the ground-state matrix element of
that fit; it is shown together with the result of the window average (black filled square) and the result
of a two-state fit to the summed correlator (black diamond).}
	\label{fig:window_fit}
\end{figure*}
	
Let $G_S\equiv \langle N | S_q| N\rangle$ denote the nucleon scalar form factor at vanishing momentum transfer.
It can be extracted from the ratio of correlation functions
	\begin{align}
		\label{eq:matelement}
                	G^\mathrm{eff}_\mathrm{S}(t,t_s)\equiv {\rm Re}\;\frac{C_3(t,t_s;{\bf 0})}{C_2(t_s;{\bf  0})} \, .
	\end{align}
Indeed, let $\Delta$ be the the energy gap between the lowest excited state and the
ground state.        
Performing the spectral decomposition in Eq.\ (\ref{eq:matelement}) and taking
the  limit of $t,(t_s-t)\gg \Delta^{-1} $, we obtain
	\begin{align}
          G^\mathrm{eff}_\mathrm{S}(t,t_s)
          &\xrightarrow{t,(t_s-t) \gg \Delta^{-1}}
		G_{\rm S}\,.
		\label{eq:asymptotic_Gs_groundstate}
	\end{align}

	We extract the ground-state contribution for each
    flavor combination of the scalar current corresponding to $\sigma_{\pi N}$, $\sigma_s$ and $\sigma_0$.
      Errors are computed using the bootstrap method on binned data with a bin size of two.
    For the conversion to physical units, we first express
dimensionful quantities in units of $t_0$ using Ref. \cite{Bruno:2016plf} (see
Tab.~\ref{tab:ensembles}) and finally use the value from \cite{FlavourLatticeAveragingGroupFLAG:2021npn} 
\begin{align}
	\sqrt{t_0}&= 0.14464(87) \, \rm{fm}
	\label{eq:t0phys}
\end{align}
to calibrate the scale.

\paragraph{\bf Excited-state analysis.}
A major obstacle to achieving reliable and precise determinations of
the ground-state matrix element is the well-known noise problem of
nucleon correlation functions \cite{Parisi:1983ae,Lepage:1989hd}.
For typical source-sink separations in current lattice calculations,
the ratio in Eq. (\ref{eq:matelement}) will be contaminated by exponentially
suppressed terms associated with resonances and multi-hadron states. 
Several approaches were
developed to have a better control over the excited-state systematics (see
\cite{Ottnad:2020qbw,Djukanovic:2021qxp} and references therein). The summation method
\cite{Maiani:1987by,Dong:1997xr,Capitani:2012gj} and multi-state fits are the most widely used among them. 
	
In the summation method, the ground-state matrix element is determined from the summed ratio
\begin{align}
        S(t_s)&=  a\sum\limits_{t= a}^{t_s-a}
	G_{\rm S}^{\text{eff}}(t,t_s)
        \stackrel{t_s \gg \Delta^{-1}}{\longrightarrow} b_1 + (t_s-a)\cdot G_{\rm S}
        \label{eq:sratio_gs_LO}  
\end{align}
by fitting $b_1$ and $ G_{\rm S}$ to $S(t_s)$.
We have extended the number of source-sink separations compared to our
analysis of the isovector vector form factor \cite{Djukanovic:2021cgp} to include smaller
source-sink
separations. This enables us to monitor the range of $t_s$ where the result from the linear
ansatz of Eq.\ (\ref{eq:sratio_gs_LO}) stabilizes. 

Rather than selecting a single fit starting at a certain value $t_s^{\rm min}$,
we follow the procedure defined in \cite{Djukanovic:2022wru} and determine $G_S$ from an average
over a range of  $t_s^{\rm min}$ values with weights
\begin{align}
	\mathit{w}(t_s^{\rm min}) &= \frac{1}{2{\cal N}}\Big[\tanh \frac{t_s^{\rm min}-t_{\text{lo}}}{\Delta
t} - \tanh \frac{t_s^{\rm min}-t_{\text{up}}}{\Delta t}\Big],
	\label{eq:weight_function}
\end{align}
with ${\cal N}$ a normalization factor.
The choice of lower ($t_{\text{lo}}$) and upper ($t_{\text{up}}$) bound
suppresses the excessive influence of excited states at small values of $t_s$
and the exponentially increasing noise at larger values, respectively.
We find the choices
\begin{align}
	t_{\text{lo}}=0.8\, \text{fm}, \quad t_{\text{up}} = 1.0\, \text{fm}
	\,\quad
	\text{and}\quad \Delta t = 0.08 \, \text{fm},
	\label{eq:choice_window}
\end{align}
to give estimates for the ground state matrix element that are more robust against statistical fluctuations
than choosing one particular value of $t_s^{\rm min}$.
Since the onset of a plateau in the extracted matrix element as a function of $t_s^{\rm min}$,
such as in the left panel of Fig.\ \ref{fig:window_fit}, 
does not entirely exclude the possibility of remnant excited-state contributions,
we also apply two further analysis methods. We note that the average using
Eq.~(\ref{eq:weight_function}) is only applied to the fit results of Eq.
(\ref{eq:sratio_gs_LO}) for different $t_s^{\rm min}$. 

As a cross check, we performed fits to the summed
correlator including the first excited state contribution,
\begin{align}	\label{eq:sratio_gs}
	S(t_s)&= \tilde b_1 + ( G_{\rm S} +\tilde{m}_{11} \,e^{ - t_s \Delta}) \,t_s \\
	&+ \frac{2 \, \tilde{m}_{10}}{\sinh(a\Delta/2)} \, e^{-{ t_s \Delta}/{2}}  \sinh \frac{(t_s -a)\Delta}{2}+\dots
\nonumber
\end{align}
where $\tilde{m}_{10}$ and $\tilde{m}_{11}$ involve matrix elements of $S_{q}$ from first excited to ground state and excited to excited state,
respectively. The excited-state contributions 
are parametrically suppressed by $\Delta\cdot t_s$.
In this case, we need priors for the energy gap $\Delta$ in order to stabilize the fits.
We choose twice the pion mass on the given ensemble as the central value and
assign
a total prior width of five percent.
Even with a prior for the energy gap $\Delta$, $\tilde{m}_{11}$ is not well constrained, and we resort to a simplified
fit ansatz excluding this term. 
\begin{figure*}[!t] \centering
	\includegraphics[width=1.0\textwidth]{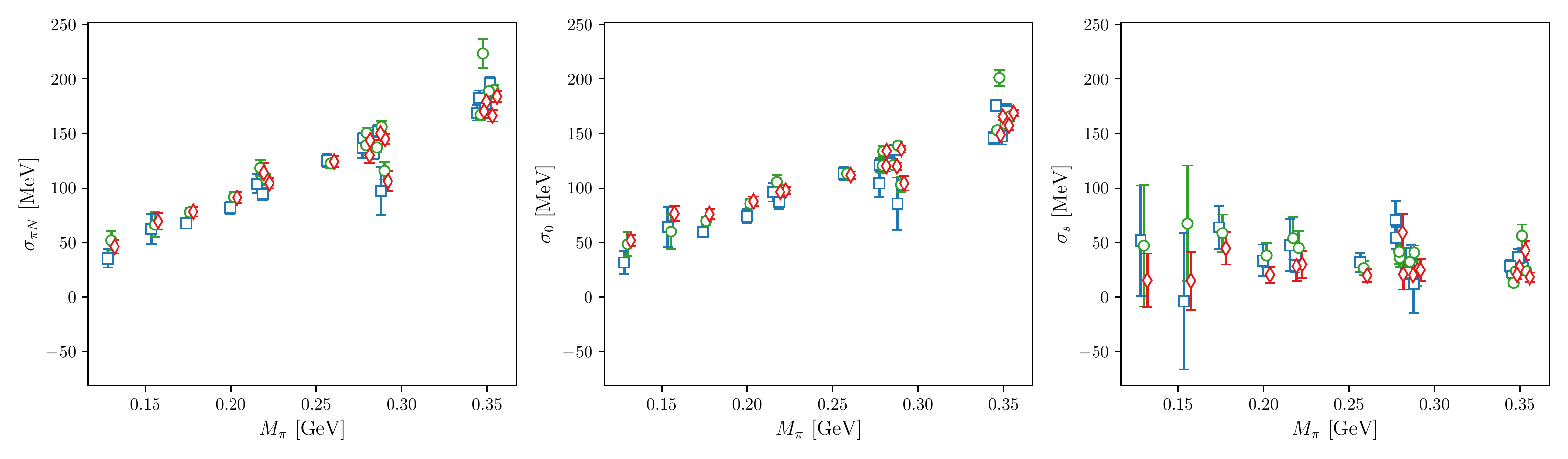}
	\caption{Comparison of the different extractions. The blue squares,
		green
		circles and red diamonds correspond to the extraction based
		on the window average of the summed correlator, the explicit
		two-state fit to the summed correlator,  and the explicit
		two-state fit to the effective form factor. 
} \label{fig:com_extractions} \end{figure*} 

In addition to the analysis of the summed correlators, we performed fits using a
two-state ansatz for the effective form factor itself. 
The fit function reads
\begin{align}
  G_{\text{S}}^{\text{eff}}(t,t_s) &= G_{\text{S}} + m_{10} \exp\left[-\Delta t  \right]
  + m_{10} \exp \left[ -\Delta (t_s-t) \right]\nonumber\\
	&+m_{11} \exp	\left[ -\Delta t_s \right]\,.
	\label{eq:twostat_fit_ansatz}
\end{align}

Similar to the analysis of the summed correlators, the gap of the
first excited state is not well constrained and we are led to using priors. For
the priors, we use the same setup as in the two-state fit to the summed
correlator.
Even though the neglected excited-state contributions in
the two-state ansatz are parametrically less
suppressed, we include all $t_s>0.8$ fm, i.e. the same value as for $t_{\rm lo}$
in Eq.\ (\ref{eq:choice_window}). Subsequently, we cut time slices at the source and
sink until a good fit is achieved. 
For $\sigma_s$ the data is too
noisy to perform two-state fits of the effective form factor, and we resort to plateau fits, where we
fit different $t_s$ and use the value that shows convergence with $t_s$.
A two-state fit applied to data at $m_\pi=200\,$MeV is illustrated in Fig.~\ref{fig:window_fit} (right panel),
along with the results of the two other methods.
Fig.~\ref{fig:com_extractions} shows a comparison of the $\sigma$ terms obtained from the different
excited-state analyses, and the results are collected in
Tab.~\ref{tab:res_sigma_piN_comp} in the appendix.

While the summation method with the averaging window fixed in units of
fm is adequate if the dominant excited-state contribution is only
weakly dependent on the pion mass, the two other analysis methods 
explicitly assume the bulk of that contribution to be associated with 
a mass gap $\Delta = {\rm O}(m_\pi)$. 
Therefore, in terms of excited-state contamination, we essentially have two
procedures, either relying on the applicability of
Eq.~(\ref{eq:sratio_gs_LO}) or, relying on assumptions about the energy gaps
through priors, applying  Eqs.~(\ref{eq:sratio_gs}) and (\ref{eq:twostat_fit_ansatz}),
where the latter are both very sensitive to the prior, but give consistent results.
In order to assess the systematics associated with the very different effects of
excited states in the two strategies, we perform the chiral and continuum
extrapolation for the window averaged summation method (fit ansatz Eq.~(\ref{eq:sratio_gs_LO})) 
and for one method using priors (fit ansatz Eq.~
(\ref{eq:twostat_fit_ansatz})), and finally model average the results with equal
weights, i.e. giving no preference to either strategy.

\paragraph{\bf Chiral and continuum extrapolation.}
The calculation of the $\sigma$ term in chiral perturbation theory (ChPT)
proceeds via
the nucleon mass using the Feynman-Hellmann theorem. The nucleon mass has been
calculated in various formulations of ChPT \cite{Borasoy:1996bx,Alarcon:2012nr,Ren:2012aj,Severt:2019sbz}
up to two-loop order \cite{Schindler:2006ha}.

Since our gauge ensembles lie on a line of constant trace of the quark
mass matrix $(2m_l+m_s)$, both the pion and the kaon mass change as $m_l$ is varied.
Moreover, to have a handle on the quantities $\sigma_0$ and $\sigma_s$,
the inclusion of the strange quark into the effective theory is mandatory. We therefore use the
result of SU(3) ChPT in the extended on-mass shell scheme (EOMS) of~\cite{Lehnhart:2004vi}.
The nucleon mass reads
\begin{align}
&	m_N = m_0 - \underbrace{(2 b_0 + 4 b_f)}_{\hat b _0} M_\pi^2 -
	\underbrace{(4 b_0 + 4 b_d - 4 b_f)}_{\hat b_1}
	M_K^2\nonumber\\
	& + {\cal F}_\pi I_{MB} (M_\pi) + {\cal F}_K I_{MB} (M_K) +{\cal F}_\eta I_{MB}
	(M_\eta),
	\label{eq:nmass}
\end{align}
with
\begin{align}
	{\cal F}_\pi &=  -\frac{3}{4}\Bigl(D^2 + 2 D F + F^2\Bigr),\\
	{\cal F}_K&= -\Bigl(\frac{5}{6} D^2 - D F + \frac{3}{2} F^2\Bigr),\\
	{\cal F}_\eta &=  -\frac{1}{2} \Bigl(\frac{1}{6} D^2 - D F + \frac{3}{2}
	F^2\Bigr),\\
	I_{MB}(M) &=  \frac{M^3}{8 F_\phi^2 m_0 \pi^2} \Biggl(
	M \log \frac{M}{m_0} \nonumber\\
	&	+\sqrt{4 -
		\frac{M^2}{m_0^2} }m_0 \arccos\Bigl(\frac{M}{2
	m_0}\Bigr)\Biggr).
	\label{eq:su3_bchpt}
\end{align}
For the $\eta$ meson mass, we assume the Gell-Mann-Okubo relation $3 M_\eta^2=4 M_K^2 -M_\pi^2$.
 We fix the values of the low-energy constants (LECs) $D=0.8$,
$F=0.46$, $F_\phi=0.108 \ {\rm GeV}$, $m_0=938.9$~MeV and fit the constants $\hat b_0$ and
$\hat b_1$. For the physical point we use the isospin-limit meson masses
$M_\pi=134.8$~MeV and $M_K=494.2$~MeV \cite{Aoki:2016frl}. 
Eq.\ (\ref{eq:nmass}) is derived with respect to the quark masses, yielding
the quark-mass dependence of the sigma terms. For the quark mass dependence of
the octet meson masses we take the leading order expression in ChPT
\cite{Gasser:1984gg}.

We treat the lattice spacing dependence of the sigma term via an additional
term\footnote{Note that the current is not $\mathcal{O}(a)$-improved.},
\begin{align}
	\sigma_{\pi N / s} \to \sigma_{\pi N /s} +   b_i\ \frac{ a}{\sqrt{t_0}}
	\, M_{\pi /K}^2 .
	\label{eq:latspac}
\end{align}
The finite-volume dependence of the nucleon mass in SU(2) is given in~\cite{Beane:2004tw},
wherefrom we derive
\begin{align}
  	\sigma_{\pi N} \to \sigma_{\pi N} + b_L \Bigl(\frac{M_\pi^3}{M_\pi L} -\frac{M_\pi^3}{2}\Bigr)\,\exp \Bigl(-M_\pi L \Bigr).
	\label{eq:parm_fv}
\end{align}
We only use the finite-volume corrections due to pion loops, as terms
$\sim \exp (- M_K L)$ are parametrically much more suppressed; thus   we omit finite-volume corrections for $\sigma_s$.
Instead of using the ChPT results for the prefactors of the finite-volume corrections,
we leave them as additional fit parameters, however we use as a loose prior the
value obtained from SU(2)~ChPT.

\begin{figure}[!t]
	\includegraphics[width=.5\textwidth]{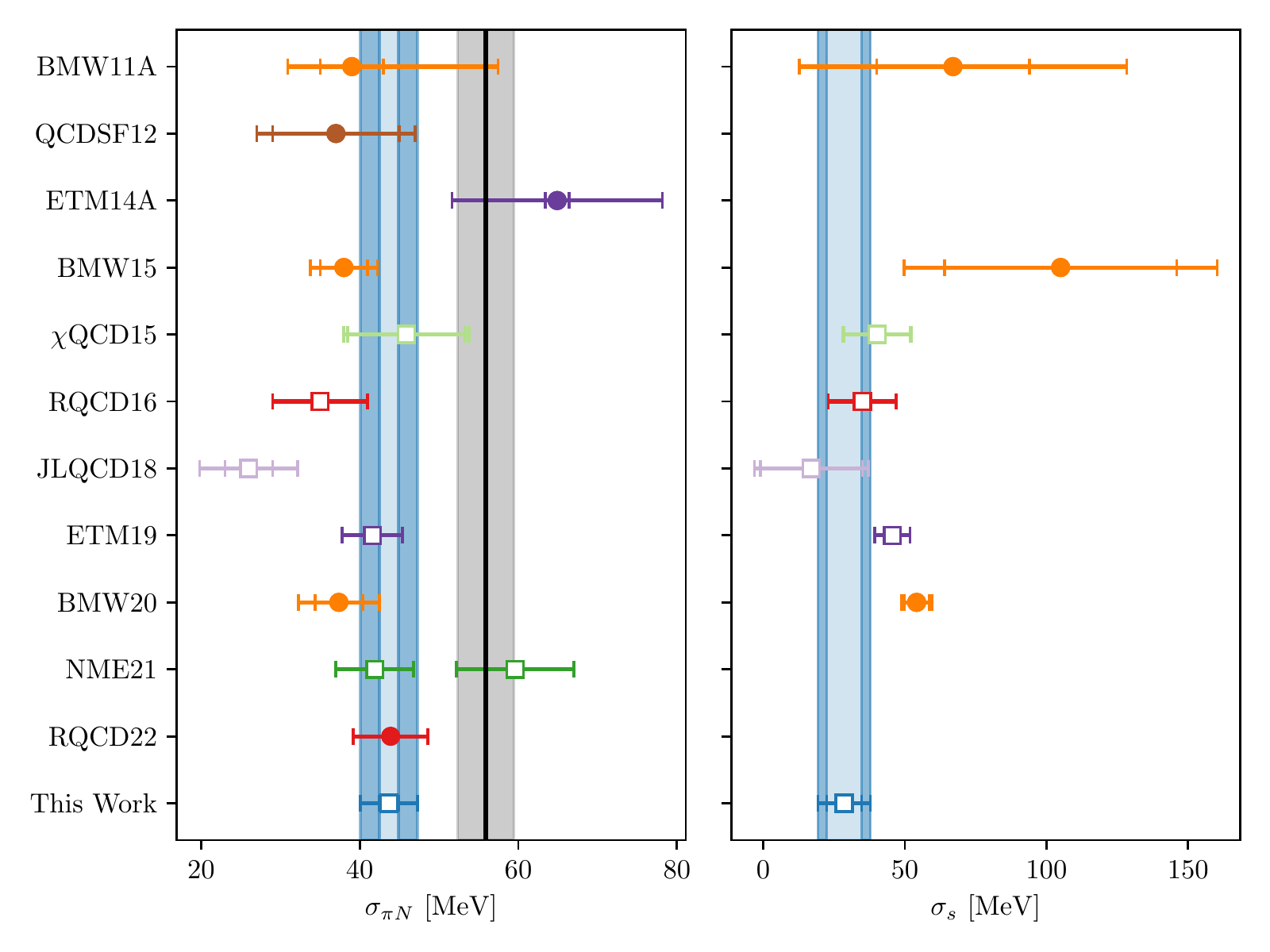}
	\caption{Comparison of our results to other lattice determinations:
		RQCD22 \cite{RQCD:2022xux},  NME21 \cite{Gupta:2021ahb}, BMW20
		\cite{Borsanyi:2020bpd},  ETM19 \cite{Alexandrou:2019brg},
		JLQCD18 \cite{Yamanaka:2018uud}, RQCD16 \cite{Bali:2016lvx},  $\chi$QCD15
		\cite{Yang:2015uis}, BMW15 \cite{Durr:2015dna}, ETM14A
		\cite{Alexandrou:2014sha}, QCDSF12
		\cite{Bali:2012qs}, BMW11A \cite{Durr:2011mp}. Filled circles
represent results extracted from the slope of the nucleon mass with
respect to the light quark mass $m_l$, and squares represent results obtained directly from the
 matrix element. The gray band corresponds to the dispersive result of
\cite{Hoferichter:2015dsa} with the correction for the isospin-limit value of
the pion mass from \cite{Hoferichter:2023ptl} applied, i.e.\ $\sigma_{\pi N}=55.9(3.5)$\,MeV.}
\label{fig:comparison}
\end{figure}

We proceed to fit $\sigma_{\pi N}$, $\sigma_s$, taking into account the correlations
among the sigma terms and lattice spacing. The fits are performed with variations in 
 the upper end of the pion mass range (220, 285 or 360\,MeV),
and  including/excluding the artifacts with respect to finite
		lattice spacing and to finite volume.
We analyze the two data sets obtained from the
 excited-states analyses  separately with
respect to the above variations, i.e.\ within each data set all variations are
averaged using an AIC weight $w_i$ given by
\begin{align}
	w_i={a_i}/{\textstyle(\sum_k a_k)}, ~~~~
  a_i&= {\exp -{\textstyle\frac{1}{2}} \Bigl [ \chi^2 + 2 n_c +2 n_f\Bigr]}
	\label{eq:aic}
\end{align}
where $n_c$ and $n_f$ denote the number of cut data points and number of fit
parameters, respectively. The weights are normalized per data set, and
 finally a flat weighting is applied between the data sets.
Using the procedure of
\cite{Borsanyi:2020mff,Djukanovic:2021cgp} we obtain as
our final estimates
\begin{subequations}
	\begin{align}
		\sigma_{\pi N} &= 43.7(1.2)(3.4) \, \rm{MeV} \\
		\sigma_{0} &= 41.3(1.2)(3.4)\, \rm{MeV} \\
		\sigma_{s} &= 28.6(6.2)(7.0) \, \rm{MeV},
	\end{align}
		\label{eq:final_numbers}
\end{subequations}
where the first and second errors correspond to the statistical and
systematic uncertainties, respectively. More details of the averaging procedure
are given in the appendix. The systematic error dominates, with the largest source of 
uncertainty  coming from the treatment of excited
states. In Fig. \ref{fig:comparison} we compare our results to those
of other lattice calculations. We note a reasonable agreement among these calculations.

\begin{table}
	\centering
	  \begin{ruledtabular}
	\begin{tabular}{ccc}
	$i:$ & window  & two-state \\ \hline
	$\sigma_{\pi N}^i$ &  42.3(2.4) MeV& 46.9(1.7) MeV \\
	$\sigma_{s}^i$ &39.6(1.9) MeV& 45.0(1.7) MeV \\
	$\sigma_0^i$ & 34.2(9.8) MeV & 24.7(6.5) MeV 
	\end{tabular}
	  \end{ruledtabular}
	  \caption{Result of the model average procedure using AIC weights defined in
		  Eq.~(\ref{eq:aic})  when applied exclusively to the data
	set denoted in the column heading. Only total errors are shown.}
	\label{tab:aic_sep}
\end{table}
\paragraph{\bf Conclusion.}
We have calculated the nucleon sigma terms $\sigma_{\pi N}$,
$\sigma_0$ and $\sigma_s$ with a full error budget concerning
excited-state contamination as well as chiral, finite-size and continuum
extrapolations. Our estimate for $\sigma_{\pi N}$ lies close to the early estimate
from $N\pi$ scattering~\cite{Gasser:1990ce}. It is compatible with
most other lattice determinations and in excellent agreement 
 with the $\sigma_{\pi N}$ determination of~\cite{RQCD:2022xux},
which uses partly the same gauge ensembles but proceeds by computing the quark-mass dependence of the nucleon mass.
For $\sigma_s$ we find a non-zero value, again compatible with most
recent lattice determinations.
Including the effects of different methods for the treatment of excited states into our error
budget, we clearly establish this to be the largest source of
systematic uncertainty. 
Analyzing the data sets from the window and two-state procedure separately,
see Tab.~\ref{tab:aic_sep}, 
we observe an upwards trend for $\sigma_{\pi N}$ when using priors similar
to~\cite{Gupta:2021ahb},  albeit not as pronounced.
Our final central value for $\sigma_{\pi N}$ lies between the two values presented in~\cite{Gupta:2021ahb},
but is much closer to that obtained without imposing tight priors on the gap $\Delta$ around values of order $m_\pi$.
A discrepancy of $2.4\,\sigma$
persists with the dispersive result of \cite{Hoferichter:2015dsa}, after applying
the correction necessary to match our definition of the pion mass in the
isospin-limit from Ref.~\cite{Hoferichter:2023ptl}
\paragraph{\bf Acknowledgments.}
We thank Marco C{\`e} for sharing his values of the PCAC masses calculated in the
context of \cite{Ce:2022eix}, and Simon Kuberski for providing improved
reweighting factors \cite{Kuberski:2023zky}
for the gauge ensembles used in our calculation.
This work was supported in part by the European Research Council (ERC)
under the European Unions Horizon 2020 research and innovation program
through Grant Agreement No. 771971-SIMDAMA and by the Deutsche
Forschungsgemeinschaft (DFG) under Grant No. HI 2048/1- 2 (Project
No. 399400745) and in the Cluster of Excellence Precision Physics,
Fundamental Interactions and Structure of Matter (PRISMA+EXC 2118/1)
funded by the DFG within the German Excellence strategy (Project ID
39083149).
Calculations for this project were partly performed on the HPC clusters
``Clover''
and ``HIMster2'' at the Helmholtz Institute Mainz, and ``Mogon 2'' at Johannes
Gutenberg- Universit\"at Mainz. The authors gratefully acknowledge the Gauss
Centre for Supercomputing e.V. (www.gauss-centre.eu) for funding this project
by providing computing time on the GCS Supercomputer systems JUQUEEN and
JUWELS at J\"ulich Supercomputing Centre (JSC) via grants NucStrucLFL, HMZ21, HMZ23 and  HMZ36
(the latter through the John von Neumann Institute for Computing (NIC)), as
well as on the GCS Supercomputer HAZELHEN at H\"ochstleistungsrechenzentrum
Stuttgart (www.hlrs.de) under project GCS-HQCD.

Our programs use the QDP++ library \cite{Edwards:2004sx}  and deflated SAP+GCR solver from the openQCD
package \cite{Luscher:2012av}, while the contractions have been explicitly checked using
\cite{Djukanovic:2016spv}. We
are grateful to our colleagues in the CLS initiative for sharing the gauge field
configurations on which this work is based.

\bibliography{sigmaterm}
\appendix

\section{Renormalization}
In the continuum, the operator $m_q \bar q q  $ is 
invariant under renormalization group transformations. However, Wilson fermions
explicitly break chiral symmetry, and this enables mixing with other quark
flavors. In the presence of chiral symmetry breaking by the regulator,
flavor-non-singlet and flavor-singlet operators represent the more adequate
basis of operators to work in. Indeed, it is straightforward to show that the
operators 
\begin{eqnarray}\label{Sigma_0,8} {\Sigma}^{(0)}&=&(2m_{{\rm q},l}+m_{{\rm q},s})(\bar
u u+\bar d d+\bar s s),\\ {\Sigma}^{(8)}&=&(m_{{\rm q}, l}-m_{{\rm q},s})(\bar u u+\bar d d-2\bar
s s), 
\end{eqnarray} 
are renormalized even in Wilson-action lattice QCD, where it is the
bare quark masses and bare scalar operators that appear on the right-hand side.
However, in order to realize $\mathcal{O}(a)$ improvement, $\mathcal{O}(a)$
counterterms are required. In
particular, terms of type $\mathcal{O}(am_q)$ and the gluonic operator $aF^2$ can be
included~\cite{Bhattacharya:2005rb}.  

The bare quark masses are related to the hopping parameters via
\eq
m_{{\rm q},f}=\frac{1}{2a}\left(\frac{1}{\kappa_{f}}- \frac{1}{\kappa_{\rm crit}} \right)\,,
\en
where $\kappa_{\rm crit}$ is the hopping parameter at which the
octet of pseudoscalar mesons becomes massless in the SU(3) symmetric theory. The
values of $\kappa_{\rm crit}$ for our action can be found in~\cite{Gerardin:2018kpy}. Using the previously defined operators $
{\Sigma}^{(0)},{\Sigma}^{(8)}$, the light- and strange-quark operators
can be reconstructed as \begin{eqnarray} m_l(\bar u u+\bar d d)&=&
	\frac{2{m}_l}{3(2{m}_l+{m}_s)}{\Sigma}^{(0)}+\frac{{m}_l}{3({m}_l-{m}_s)}{\Sigma}^{(8)}~~~~~\\
m_s\bar s s&=&
\frac{{m}_s}{3(2{m}_l+{m}_s)}{\Sigma}^{(0)}-\frac{{m}_s}{3({m}_l-{m}_s)}{\Sigma}^{(8)}.
~~~~~
\end{eqnarray} An important observation is that the ratios of quark masses
appearing in these expressions can be evaluated using the PCAC (partially conserved axial current) quark masses
(see e.g.\ \cite{Gerardin:2018kpy} for their definition, including ${\cal O}(a)$ improvement).
Proceeding in this way by-passes the use of the finite renormalization
factor $\rrm$, which parametrizes the difference in renormalization of
the SU(3)$_{\rm f}$ octet and singlet quark mass combinations~\cite{Bhattacharya:2005rb}.
We note that for QCD actions with an exact chiral symmetry $\rrm=1$~\cite{Takeda_2011}.

\subsection{O($a$) improvement of $\Sigma^{(0)}$ and $\Sigma^{(8)}$}

In the following we estimate the size of the $\mathcal{O}(a)$ corrections for
the flavor-singlet and non-singlet scalar operator in $N_{\rm f}=2+1$ flavor QCD.
First, we recall that the renormalization and improvement pattern of a non-singlet combination of quark masses reads~\cite{Bhattacharya:2005rb}
\be\la{eq:mlms}
\hat m_l - \hat m_s = \zm(\mql-\mqs) \Big[ 1+ 3\bbarm a\mqav + a b_{\rm m} (\mql+\mqs)\Big],
\ee
while the singlet combination renormalizes as 
\ba\la{eq:TrM}
&& \widehat{{\rm Tr}(M)} \equiv (2m_l+m_s)_{R,I}
\\ &&= \zm \,\rrm \Big[ (1+a\dbarm 3 \mqav) 3\mqav + a\dm (2\mql^2+\mqs^2)\Big].
\nonumber
\ea
We follow the notation of~\cite{Bhattacharya:2005rb}, denoting by a hat 
an operator or a parameter of the theory that has been renormalized and $\mathcal{O}(a)$ improved. 
\subsection{The octet scalar operators}
Let $S_{f',f}\equiv \bar\psi_{f}\psi_{f'}$ and $\lambda^a$ denote a Gell-Mann matrix.
We then define $S^a \equiv {\rm Tr}\{\lambda^a S \} = \bar\psi {\lambda^a}\psi$ to be the octet of scalar currents,
and  ${\rm Tr\,} S =\bar\psi  \psi$ to be the flavor-singlet current.
The octet of scalar currents has no additive improvement term in the massless
limit.
Thus the renormalization and improvement pattern of the local discretization of the 
two neutral octet combinations reads~\cite{Bhattacharya:2005rb}
\ba\label{eq:S3}
\widehat S^3 &=& \zs\; (1+ 3\bbars\; a\mqav + \bs\;a\mql) \;S^{3},
\\
\label{eq:S8}
\widehat S^8 &=& \zs\; \bigg[ \Big(1+ 3 \bbars\; a\mqav + \frac{\bs}{3}\; a (\mql+2\mqs\Big) \, S^{8} 
\nonumber    \\ && \qquad ~       +\, ({\textstyle\frac{1}{3}}\bs+\fs)\;
\frac{2}{\sqrt{3}}a(\mql-\mqs) \;{\rm Tr\,}S\bigg]
\ea
where $\zs = 1/\zm$, and the improvement coefficients are not independent~\cite{Bhattacharya:2005rb},
\be
\bs=-2 b_{\rm m}, \quad \bbars = - \bbarm, \quad 3 f_s = 2(b_{\rm m}-\dm).
\ee
Taking into account these relations, we obtain
\ba
\widehat\Sigma^{(8)} &\equiv& (\hat m_l - \hat m_s) (\bar u u + \bar d d - 2\bar s s)_{R,I}
\nonumber \\ &=& (\mql - \mqs) \Big[ \nonumber\\
	&&(\bar u u + \bar d d) \Big( 1+ a(\mql -
	\mqs) {\txts\frac{1}{3}} (b_{\rm m}-4\dm)\Big)
\nonumber\\ &&   - 2\bar s s \Big( 1+ a(\mql- \mqs){\txts\frac{1}{3}} (b_{\rm m}+2\dm)\Big) \Big].
\ea
The difference $\dm -b_{\rm m}$ is a sea-quark effect that we will neglect in the following. Furthermore, $\bs = -2b_{\rm m}$ has been determined
on CLS ensembles in~\cite{Korcyl:2016ugy}. We note that in perturbation theory, $b_{\rm m}=-\frac{1}{2}+ {\rm O}(g_0^2)$.
At $\beta=3.55$ for instance, Ref.\ \cite{Korcyl:2016ugy} finds $b_{\rm m} =-0.835$. Since on ensemble D200 
$a(\mql - \mqs) = \frac{1}{2\kappa_l} - \frac{1}{2\kappa_s}\simeq -0.0160$, we arrive at the estimates
\ba\la{eq:est1}
 a(\mql - \mqs) {\txts\frac{1}{3}} (b_{\rm m}-4\dm) &\approx&  - 0.0134,
\\
a(\mql- \mqs){\txts\frac{1}{3}} (b_{\rm m}+2\dm)  &\approx&  + 0.0134.
\la{eq:est2}
\ea
The O($a$) corrections are thus on the order of a few percent. 
They reduce slightly the weight of the light quarks and increase the magnitude of the weight of the strange quark.

\subsection{The flavor-singlet scalar operator}

\begin{figure*}[!t]
	\centering
	\includegraphics[width=.9\textwidth]{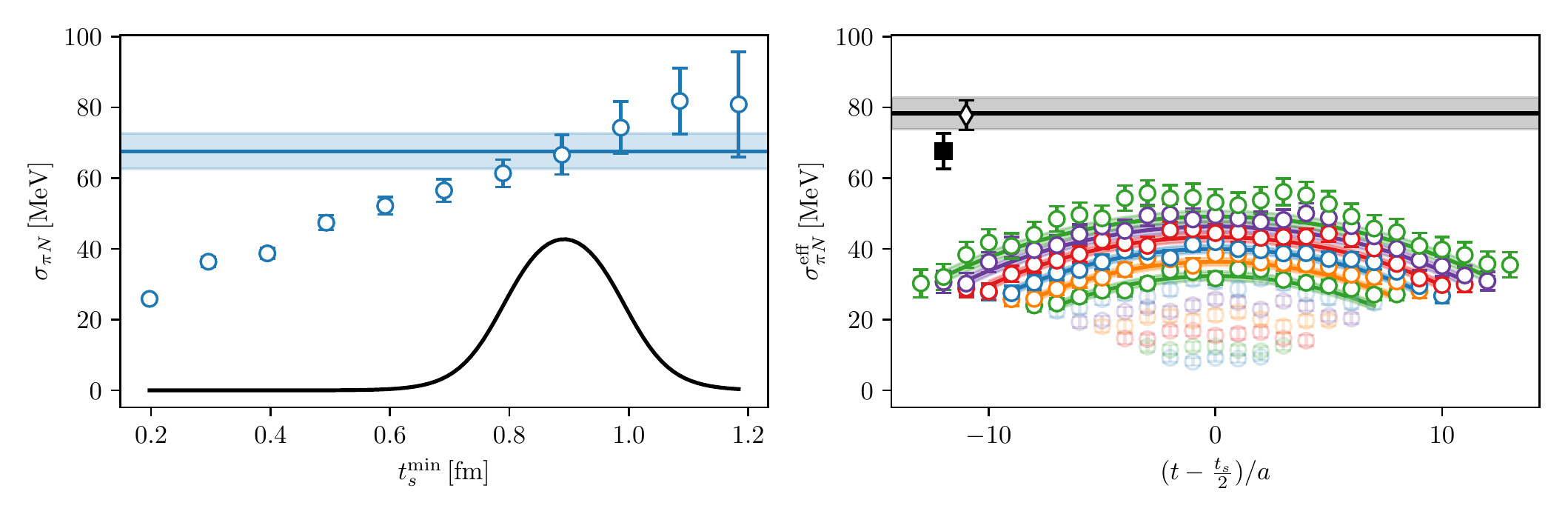}
\caption{Left: Results of linear fits to the summed
        correlator on ensemble E300 with the starting time slice given on the
	$x$-axis. The blue
        shaded area is the weighted average using the weight function of Eq.
(\ref{eq:weight_function}) shown as a black line in the bottom of the plot, for
the particular choice of parameters from Eq. (\ref{eq:choice_window}). Right:
Fit result of an explicit two-state fit to the effective form
factor. The gray band denotes the result for the ground-state matrix element of
that fit, together with the result of the window average (black filled square) and the result
of a two-state fit to the summed correlator (black diamond).}
	\label{fig:E300_eff_sum}
\end{figure*}
The improvement of the scalar operator
 \begin{align}
 {\rm Tr\,} S \equiv \bar u u+ \bar d d + \bar s s
 \end{align}
in the SU(3) chiral limit is  given by 
\be
{\rm Tr}(S)^I = {\rm Tr}(S) + a\gs \,{\rm Tr}(F_{\mu\nu}F_{\mu\nu}),
\ee
where the gluonic action is given by 
\begin{align}
	S_G = \frac{a^4}{2g_0^2} \sum_x  {\rm Tr}(F_{\mu\nu}F_{\mu\nu}).
\end{align}
The renormalized, improved scalar operator then reads
\be\la{eq:TrS}
\widehat{{\rm Tr}(S)} = \zs \,\rs \Big[ (1+a\dbars 3\mqav) \,{\rm Tr}(S)^I + a\ds\,\widehat{{\rm Tr}(MS)} \Big].
\ee
Note that $MS = {\rm diag}(\mql \bar u u, \; \mql \bar d d,\; \mqs \bar s s)$.

Again, using $\zs = 1/\zm$ and $\rs=1/\rrm$, as well as the
 relations among the improvement coefficients~\cite{Bhattacharya:2005rb}
\ba
\ds &=& -(2b_{\rm m}+ 3 \bbarm),
\\
\dbars  &=& \frac{2}{3} (b_{\rm m} - \dm) + \bbarm  - 2\dbarm,
\ea
one obtains
\ba
\widehat{\Sigma^{(0)}} &\equiv& \widehat{{\rm Tr}(M)} \,\widehat{{\rm Tr}(S)}
\nonumber\\
&=& (\bar u u + \bar d d)\Big[ 3\mqav 
+  b_{\rm m} 2a\mqav (\mqs-\mql) 
\nonumber\\ && + a(\mqs^2 -8\mql \mqs - 2\mql^2) \frac{\dm}{3}
\nonumber\\ &&  + \bbarm a (\mqs^2 +\mqs \mql - 2\mql^2) \nonumber\\&&- \dbarm a(2\mql+\mqs)^2 \Big]
\nonumber\\ && +  \bar s s \Big[ 3\mqav 
+ 4a(\mql-\mqs) \mqav b_{\rm m}  
\nonumber\\ &&  + a(\mqs^2 -8\mql \mqs - 2\mql^2) \frac{\dm}{3}
\nonumber\\ &&   -2 \bbarm a (\mqs^2 +\mqs \mql -2\mql^2)\nonumber\\&& - \dbarm a(2\mql+\mqs)^2\Big]
\nonumber \\ && + 3a\mqav\,\gs\,{\rm Tr}(F_{\mu\nu}F_{\mu\nu}).
\label{eq:singlet_oa}
\ea
In order to estimate the size of the $\mathcal{O}(a)$ correction we take $\dm\simeq b_{\rm m}$ from~\cite{Korcyl:2016ugy}
and $\bbarm \simeq 0 \simeq \dbarm$. If $\mql\ll\mqs$, then 
the O($a$) corrections in the square brackets have the same relative size 
as in the octet case, Eq.\ (\ref{eq:est1}--\ref{eq:est2}).

As for $\gs$, we note the relation~\cite{Bhattacharya:2005rb} 
\be
\gs = \frac{1}{2g_0^2} \bg = 0.018000(3) \frac{N_{\rm f}}{3}\, + {\rm O}(g_0^2),
\ee
where we used the one-loop result of~\cite{Sint:1995ch} for $\bg$.
The trace anomaly in the nucleon at rest yields
(see for instance~\cite{Ji:1994av}; we use the non-relativistic normalization of the nucleon state at rest)
\begin{align}
\langle N|T_{\mu\mu}^{\rm g}|N\rangle  \simeq 0.800\,{\rm GeV}.
\end{align}
The trace anomaly is related to the improvement term via
\begin{align}
  T_{\mu\mu}^{\rm g}
  = \frac{\beta(g_0)}{g_0^3}  \,{\rm Tr}(F_{\mu\nu}F_{\mu\nu})
\end{align}
with $\beta(g) = -b_0g^3 + \dots$,
\begin{align}
b_0 &= \frac{1}{4\pi^2}(\frac{11}{3}N_c - \frac{2}{3} N_{\rm f})=0.228.
\end{align}

Hence the last term in Eq.~(\ref{eq:singlet_oa}) is of the order of $-2$ MeV for
$a=(3 {\rm GeV})^{-1}$.
For $\sigma_{\pi N}$ there is a suppression by $\frac{2}{3} \frac{m_l}{2m_l+m_s}\simeq 0.025$.
Thus we expect the lattice artifacts due to the gluonic operator to be of order
$-0.05$MeV in $\sigma_{\pi N}$.
This is certainly negligible, even if the perturbative estimate of $\gs$ was too small by an order of magnitude.
\section{Correlator Analysis Details}
In our calculation of the sigma terms the statistical precision of the
correlator is restricted by two factors, the signal-to-noise of the disconnected
contribution and the occurrence of exceptional configurations which produces outliers. The signal-to-noise problem of the
disconnected part is
exacerbated at small pion mass, since the absolute contribution of the
connected part decreases. In the following we summarize our strategy to deal
with these two problems.
\subsection{Disconnected Part}
We observe that the  connected part for a given statistics is far more precise
than the disconnected. The disconnected part consists of the loop contribution and
the two-point function. For the quark loops, we have exhausted the number of sources per
configuration, for which error scaling still holds, on most of our ensembles.
Consequently we concentrate on the 2-point function, i.e. improving the signal of the disconnected part by using additional sources for
the 2-point functions.
For the connected part we keep a matched setup between 2- and 3-point
functions, as the correlation plays an important role. In Fig.\ref{fig:E300_eff_sum} we show a
comparison of the excited state analysis between summation and explicit
two-state ans\"atze for E300. Here the signal for the effective form factor has
been improved via additional two-point functions. However, even after including
additional sources statistical fluctuations are still clearly visible. 
% Remove if single document

The different strategies are explained in the main text, for the window average
the weight function reads
\begin{align}
	\mathit{w}(t_s) &= \frac{1}{2{\cal N}}\Big[\tanh \frac{t_s-t_{\text{lo}}}{\Delta
t} - \tanh \frac{t_s-t_{\text{up}}}{\Delta t}\Big],
	\label{eq:weight_function_apx}
\end{align}
with the particular choice
\begin{align}
	t_{\text{lo}}=0.8\, \text{fm}, \quad t_{\text{up}} = 1.0\, \text{fm}
	\,\quad
	\text{and}\quad \Delta t = 0.08 \, \text{fm}.
	\label{eq:choice_window_apx}
\end{align}

\begin{table}[!t]
	\begin{tabular}{ccc}\hline\hline
		Ensemble &  Exact Sources & Sloppy Sources\\ 
		{\bf{C101}} & 7999 & 239405\\ %  
		{\bf{H102}} & 14033 & 413083 \\ % 
		{\bf{H105}} & 25660 & 249059 \\ % 
		{\bf{N101}} & 3170 & 400995 \\  %
	{\bf{D200}} &  9000 & 271802 \\ %
	{\bf{E250}} & 6400 & 204800\\   %  
	{\bf{N200}} & 13696 & 406966\\  %
	{\bf{N203}} &  6175 & 178587 \\ %
	S201 & 4181 & 96280 \\   %
	{\bf{J303}} & 3219 & 145872\\   %
	{\bf{N302}} & 8797 & 237034 \\  %
	{\bf{E300}} & 1138 & 163872 \\  % 
	N451 & 8088 & 129408 \\  % 
	D452 & 8000 & 128000 \\  % 
	D450 & 4000 & 64000 \\   % 
	S400 & 5742 & 67543 \\   % 
\end{tabular}
\caption{Number of exact and sloppy sources for the calculation of the two-point
function, which enters the estimate of the disconnected contribution. Ensembles
in bold, number of sources is increased with respect to the statistics for the
connected three-point function.}
\label{tab:2pt_stat}
\end{table}
\subsection{Outliers}
In contrast to other observables in our previous analysis, we observe a small
number of  configurations for which the effective form
factor of the scalar operator exceeds the ensemble average by a huge
amount, which potentially spoils the correct estimation of the error.
Removing the measurements
on a (negligibly small) fraction of the configurations considerably improves the
error estimate.

\begin{figure*}[!t]
	\centering
	\includegraphics[width=.9\textwidth]{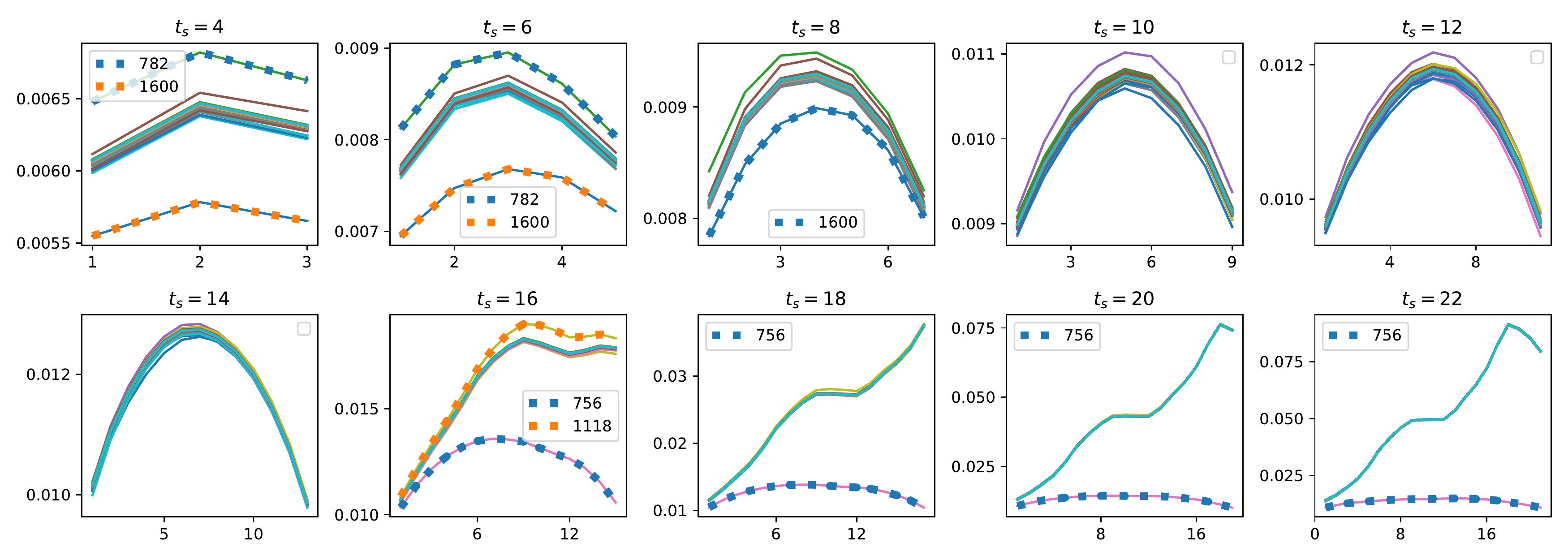}
	\caption{Jackknife distribution of the connected part for the
	$\sigma$-term for all source-sink-separations. Dashed are the flagged
configurations by the analysis described in the main text. For D200 a total of 4
configurations are flagged out of 2000.}
	\label{fig:D200_outlier_connected}
\end{figure*}
We may assume the sampling distribution, either Jackknife
or bootstrap, in the limit of a large number of measurements to be normally
distributed. However on some of the ensembles the observed distribution deviates
strongly from a normal distribution. The most
prominent example is ensemble D200, where we identify one configuration to be
the root cause of the gross overestimation of errors, i.e. an outlier in our
analysis. It is well known that the mean is not a robust estimator with respect
to outliers and may be highly affected by the existence of extreme values on the correlator level.
We try to identify the extreme values on a per-configuration basis performing
first a Jackknife analysis,
where the sampling distribution should be approximately Gaussian. We essentially
look for extreme deviations from the central location of the sampling distribution
on each time slice and for every source-sink separation for the effective form
factor corresponding to the connected and disconnected part of the
sigma terms, respectively. Whenever we find a
value that is more than $\sim \, 6\sigma$ away, we flag the Jackknife sample, i.e. configuration, and remove it
from the subsequent analysis, see Fig.~\ref{fig:D200_outlier_connected}. For
estimating the central location of
the distribution and its standard deviation we use the median and median
absolute deviation as robust replacements for the mean and standard deviation.
For the latter we apply the usual correction factor $\Phi^{-1}(3/4)=0.67449$ to
make contact with the standard deviation of a normal distribution.
We find this procedure correctly identifies all problematic results, that either are not
symmetric with respect to $t_s/2$ and/or are very far away from the center
of observations. Data from the latter category may also come from a sampling
distribution that has a longer tail than the normal distribution. We
therefore apply a very
loose cut using $\sim \, 6\sigma$, i.e. the number of flagged configurations is kept to a
minimum, so as to not distort the empirical distribution.
\begin{table}
\begin{tabular}[t]{cc|cc}
	\hline\hline
	Ensemble & \# flagged configs & Ensemble & \#
	flagged configs\\\hline
	C101 & 9, 2 & H102 & 1, 0\\
	N101 & 5, 1 & H105 & 4, 0\\
	E250 & 6, 0 & D200 & 4, 1\\
	N200 & 2, 1 & N203 & 0, 0\\
	S201 & 1, 0 & E300 & 0, 0\\
	J303 & 0, 0 & N302 & 0, 0\\
	D452 & 6, 4 & D450 & 3, 0\\
	N451 & 0, 0 & S400 & 0, 0\\
\end{tabular}
\caption{Number of flagged configurations for each ensemble. The first number
refers to the connected part, while the second concerns the disconnected.}
\end{table}
The number of flagged configurations is generally well below 1\%, except for
E250 with 1.5 \%, which also has the smallest number of configurations amongst
the ensembles analyzed.
\section{Results for the sigma terms}
The results for the three determinations described in the main text for all
ensembles  are collected in Tab.~\ref{tab:res_sigma_piN_comp}.
\begin{table*}[!t]
        \begin{ruledtabular}
                \begin{tabular}{c|ccc|ccc|ccc} 
                        ID & $\sigma_{\pi N}^{\rm window}$ & $\sigma_{\pi
                        N}^{\rm sum \ two-state}$ &$\sigma_{\pi
                        N}^{\rm two-state}$ & $\sigma_{0}^{\rm window}$ &
                        $\sigma_{0}^{\rm sum \ two-state}$ &$\sigma_{0}^{\rm
                        two-state}$ & 
                        $\sigma_{s}^{\rm window}$ &
                        $\sigma_{s}^{\rm sum \ two-state}$ &$\sigma_{s}^{\rm
                        two-state}$ \\ \hline
H102 & 195.9(5.6)	    &189.8(4.9)	&183.9(5.1)	&170.6(6.9)	&166.0(4.5)	&168.8(3.0)	&27.1(4.9)	&23.8(3.8)	&18.1(4.1) \\
N101 & 145.2(5.4)	    &150.5(5.0)	&143.7(4.9)	&121.0(6.2)	&133.6(4.7)	&133.9(3.7)	&54(10)	&36.1(8.6)	&21(14) \\
H105 & 136.8(9.7)	    &139.3(6.7)	&129.9(7.0)	&105(13)	&120.4(6.4)	&120.0(5.0)	&71(17)	&42(11)	&59(17) \\
C101 & 94.3(5.6)	    &108.0(5.1)	&104.5(4.8)	&87.0(6.5)	&98.2(5.2)	&97.7(3.7)	&39(17)	&45(15)	&30(12) \\
S400 & 177.9(9.4)	    &188.7(6.1)	&166.3(5.4)	&147.5(7.4)	&158.7(3.2)	&157.2(3.9)	&36.4(7.9)	&56(11)	&42.6(8.9) \\
N451 & 152.1(5.6)	    &156.1(5.0)	&145.1(4.4)	&135.0(4.1)	&138.9(3.7)	&135.5(3.0)	&40.1(7.8)	&40.8(6.4)	&24.8(4.0) \\
D450 & 103.8(8.9)	    &118.3(7.5)	&113.9(8.9)	&96.2(8.6)	&105.6(6.6)	&96.4(4.0)	&47(24)	&54(19)	&28(13) \\
D452 & 62(14)	    &66(12)	&69.6(7.5)	&64(18)	&60(16)	&76.6(6.8)	&-4(62)	&67(53)	&15(27) \\
N203 & 182.7(6.7)	    &223(13)	&179.2(5.4)	&175.7(4.3)	&201.1(7.5)	&165.5(3.1)	&22.5(5.0)	&23.8(3.7)	&27.2(6.9) \\
S201 & 97(22)	    &115.7(7.9)	&106.3(9.0)	&85(24)	&103.1(6.9)	&104.4(6.8)	&12(27)	&23(12)	&24.8(9.9) \\
N200 & 131.6(5.4)	    &137.3(3.8)	&150.1(6.0)	&122.7(5.6)	&121.0(3.2)	&119.9(3.3)	&22.2(6.0)	&32.4(4.4)	&20.1(4.4) \\
D200 & 81.8(5.8)	    &91.6(4.3)	&91.0(5.0)	&74.2(6.5)	&85.8(4.2)	&87.6(4.3)	&33(15)	&38(11)	&20.2(7.5) \\
E250 & 35.44(8.39)	    &51.77(8.77)	&46.19(6.18)	&32(11)	&48(11)	&51.8(5.3)	&52(51)	&47(56)	&15(25) \\
N302 & 168.8(7.1)	    &167.2(4.6)	&170.4(6.2)	&145.8(5.8)	&152.9(3.0)	&149.0(3.3)	&28.3(5.7)	&13.0(3.4)	&20.3(3.9) \\
J303 & 124.8(6.1)	    &122.4(4.3)	&124.1(4.9)	&113.3(5.6)	&113.2(3.4)	&111.8(3.3)	&31.8(8.9)	&26.5(6.5)	&19.6(6.2) \\
E300 & 67.6(5.0)	    &77.8(4.2)	&78.3(4.5)	&59.5(4.4)	&69.8(3.4)	&76.1(4.7)	&64(20)	&58(17)	&45(15) \\
\end{tabular}
        \end{ruledtabular}
        \caption{Results for the sigma terms on every ensemble in MeV, where
                window, sum two-state and two-state, refer to the window average
                of the summed correlator, the two-state fit to the summed
                correlator and the direct two-state fit to the correlator,
        respectively.}
        \label{tab:res_sigma_piN_comp}
\end{table*}

The conversion to physical units uses the ratios $\frac{t_0}{a^2}$ from
\cite{Bruno:2016plf} and 
\begin{align}
        \sqrt{t_0}&= 0.14464(87) \, \rm{fm}
        \label{eq:t0physApdx}
\end{align}
at the physical point from \cite{ FlavourLatticeAveragingGroupFLAG:2021npn}. The error estimate is based on Bootstrap procedure with a sample size of 5000.

\section{Fits and model average}

\begin{figure*}[!t]
	\includegraphics[width=.8\textwidth]{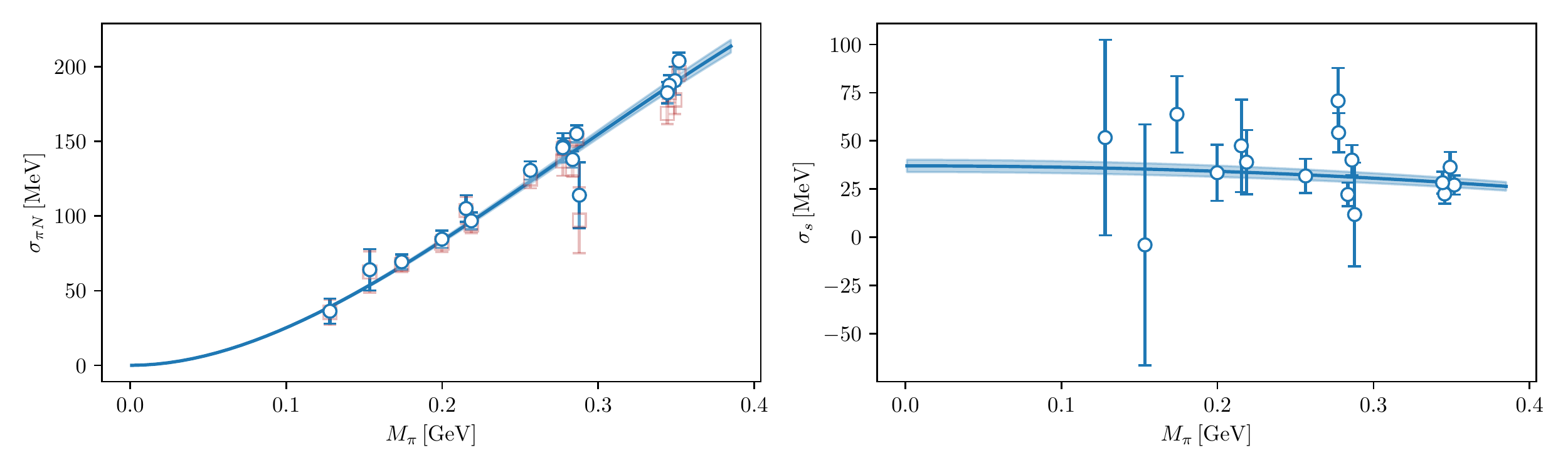}
	\caption{Simultaneous fit of the window averaged data for $\sigma_{\pi
	N}$, $\sigma_s$. In this variation we include corrections due
	to finite volume, using all available pion masses. We correct the
central value for the fitted finite volume correction only.} 
\label{fig:sumwindow_chpt_fit}
\end{figure*}

In Fig.~\ref{fig:sumwindow_chpt_fit} we show one particular fit for the summation window
averaged data based on  the
SU(3) formula for the nucleon mass Eq.~(\ref{eq:nmass}) without any cut in the pion mass including finite
size effects. The data have been corrected for finite volume effects only, while the fit
is at physical kaon mass.

We derive the expression for the sigma terms from the nucleon mass
\begin{align}
&	m_N = m_0 - \underbrace{(2 b_0 + 4 b_f)}_{\hat b _0} M_\pi^2 -
	\underbrace{(4 b_0 + 4 b_d - 4 b_f)}_{\hat b_1}
	M_K^2\nonumber\\
	& + {\cal F}_\pi I_{MB} (M_\pi) + {\cal F}_K I_{MB} (M_K) +{\cal F}_\eta I_{MB}
	(M_\eta),
	\label{eq:nmass_apx}
\end{align}
with
\begin{align}
	{\cal F}_\pi &=  -\frac{3}{4}\Bigl(D^2 + 2 D F + F^2\Bigr),\\
	{\cal F}_K&= -\Bigl(\frac{5}{6} D^2 - D F + \frac{3}{2} F^2\Bigr),\\
	{\cal F}_\eta &=  -\frac{1}{2} \Bigl(\frac{1}{6} D^2 - D F + \frac{3}{2}
	F^2\Bigr),\\
	I_{MB}(M) &=  \frac{M^3}{8 F_\phi^2 m_0 \pi^2} \Biggl(
	M \log \frac{M}{m_0} \nonumber\\
	&	+\sqrt{4 -
		\frac{M^2}{m_0^2} }m_0 \arccos\Bigl(\frac{M}{2
	m_0}\Bigr)\Biggr).
	\label{eq:su3_bchpt_apx}
\end{align}
Using the
lowest order ChPT expression for the quark mass dependence of the meson masses,
the sigma terms read
\begin{subequations}
\begin{align}
	\sigma_{\pi N}&= \frac{M_\pi}{2}  \frac{\partial m_N}{\partial M_\pi} +
	\frac{M_\pi^2}{4 M_K} \frac{\partial m_N}{\partial M_K} 
	+\frac{M_\pi^2}{6 M_\eta } \frac{\partial m_N}{\partial M_\eta}\\
	\sigma_s&=\frac{2 M_K^2 - M_\pi^2}{4 M_K} \frac{\partial m_N}{\partial
	M_K} 
	+ \frac{2 M_K^2 - M_\pi^2}{3 M_\eta} \frac{\partial m_N}{\partial
	M_\eta},\\
	\sigma_0 &=\sigma_{\pi N} - \frac{2 M_\pi^2}{2M_K^2 - M_\pi^2} \sigma
	_s.
	\label{eq:sigfit_apx}
\end{align}
\end{subequations}

\begin{figure*}[!t]
	\includegraphics[width=1\textwidth]{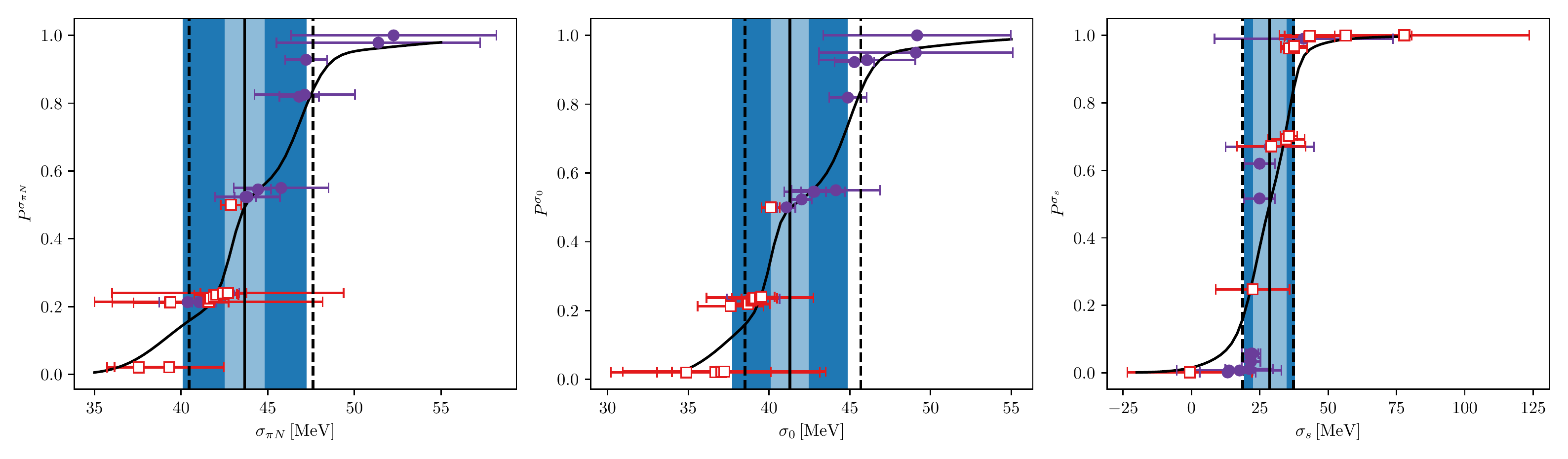}
	\caption{Cumulative distribution function of
		Eq.~(\ref{eq:cdf_aic_average_apx}) for
		all variations of the fits
		to the 
	sigma-terms. The red squares and  purple dots denote fits based on summation
window and two-state data, respectively.}
	\label{fig:cdf}
\end{figure*}

The fits are performed simultaneously to $\sigma_{\pi N}$, $\sigma_s$ and $m_N$,
where we include the correlations among 
the sigma terms.
We perform variations of these fits, i.e. three cuts in the pion mass,
including/excluding lattice spacing, including/excluding finite volume and
including both lattice spacing and finite volume corrections. The strictest pion
mass cut is such that enough data points remain to perform the fit
using all values of the lattice spacing. We treat the data
subset selection  problem using the ``perfect model'' method of Ref.~\cite{Neil:2023pgt}.  In total we thus
have 12 variations on three data sets. Instead of choosing a particular fit we
perform model averages over the 36 fits using their AIC weights.  As described in the main text,
only two data sets enter the final analysis. The weights are
normalized first on each data set, and subsequently averaged using flat weights,
i.e. with a factor $1/2$. From these weights we build a cumulative distribution
function (see Fig.~\ref{fig:cdf}) following Ref.~\cite{Borsanyi:2020bpd}
\begin{align}
	P^x(y)=\int \limits_{-\infty} ^ y \sum\limits_{i}^n w_i
	\mathcal{N}(y';x_i,\sigma_i^2) dy'
	\label{eq:cdf_aic_average_apx}
\end{align}
We estimate the central value and the total error of the average, using  the
median and the difference between the 1-$\sigma$ percentiles of $P^x$. 
For the separation into statistical and systematic errors we assume that
        \begin{align}
                \sigma_{\text{stat}}^2+\sigma_{\text{sys}}^2 =
                \sigma_{\text{total}}^2,
        \end{align}
        and that a scaling of the individual Bootstrap errors with an arbitrary
        constant is expected to affect $\sigma_{\text{stat}}^2$ exclusively. In
Fig.~\ref{fig:cdf} we show the CDF for all three quantities, note that
$\sigma_0$ is not fitted. The blue shaded area is the symmetric error from the
percentiles centered around the median. These coincide rather
well with the 1-$\sigma$ percentiles of the actual distribution.
The breakup into systematic and statistical error uses the fact that the
systematic error does not change if all errors are inflated by some arbitrary
factor (see Ref.~\cite{Borsanyi:2020mff} for more details).

In Tab.~\ref{tab:extrap_res} we collect the results for the different variations and the
corresponding weights in the averaging procedure $w_i$. We note that we
performed the averaging procedure with all quantities expressed in units of
$t_0$, applying the calibration in the end to convert to physical units. 
We see that in general the fit quality for all variations is acceptable. The
penalty terms in the AIC weights prefer variations with more data and fewer fit
parameters. That is visible for the window data, where most of the weight is on
the fits using all available pion masses and including a finite volume correction.
On the other hand, the two-state data prefers fits with stricter cuts in the pion
mass, and again finite volume corrections. 
When
performing the analysis separately for the summation-window and two-state data
for the sigma terms we obtain
\begin{align}
	\sigma_{\pi N}^{\text{window}}&= 42.3(2.4)\, \text{MeV}\\
	\sigma_{\pi N} ^{\text{two-state} }&= 46.9(1.7)\, \text{MeV}\\
	\sigma_{0}^{\text{window}}&= 39.6(1.9)\, \text{MeV}\\
	\sigma_{0} ^{\text{two-state} }&= 45.0(1.7)\, \text{MeV}\\
	\sigma_{s}^{\text{window}}&= 34.2(9.8)\, \text{MeV}\\
	\sigma_{s} ^{\text{two-state} }&= 24.7(6.5)\, \text{MeV},
\end{align}
where only the total error is given.
All values are within $1-\sigma$ of our best estimate, as can be seen in
Fig.~\ref{fig:cdf}, where the bulk of points is covered by the total errors of
our best estimate.

\begin{table*}[!t]
	\begin{tabular}{lccccc}
		\hline\hline
		Variation & $\sigma_{\pi N}$ [MeV] & $\sigma_0$ [MeV] &
		$\sigma_s$ [MeV] &
		$\chi^2$(dof) & weight in \%\\\hline
		$M_\pi < 220$ MeV & 42.04(1.27) & 38.70(1.35) & 43.18(9.20) & 4.0(10)  & 1 \\
$M_\pi < 285$ MeV & 41.89(67) & 38.98(69) & 37.56(4.74) & 20.5(18)  & 0 \\
no cut in $M_\pi$ & 41.67(44) & 38.91(41) & 35.62(3.09) & 42.9(30)  & 1 \\
$M_\pi < 220$ MeV+ $\mathcal{O}(a)$ & 41.58(6.58) & 37.23(6.28) & 56.36(24.19) & 3.5(8)  & 0 \\
$M_\pi < 285$ MeV+ $\mathcal{O}(a)$ & 39.31(3.15) & 37.05(3.06) & 29.24(12.55) & 19.6(16)  & 0 \\
no cut in $M_\pi$+ $\mathcal{O}(a)$ & 37.55(1.82) & 34.87(1.80) & 34.68(6.69) & 37.5(28)  & 2 \\
$M_\pi < 220$ MeV+ $\mathcal{O}(e^{-mL})$ & 42.45(1.33) & 39.10(1.40) & 43.26(9.20) & 3.8(9)  & 0 \\
$M_\pi < 285$ MeV+ $\mathcal{O}(e^{-mL})$ & 42.43(79) & 39.53(81) & 37.52(4.74) & 19.9(17)  & 0 \\
no cut in $M_\pi$+ $\mathcal{O}(e^{-mL})$ & 42.87(59) & 40.11(57) & 35.78(3.09) & 34.4(29)  & 26 \\
$M_\pi < 220$ MeV+ $\mathcal{O}(a)$+ $\mathcal{O}(e^{-mL})$ & 42.69(6.68) & 36.67(6.47) & 77.88(45.65) & 3.2(7)  & 0 \\
$M_\pi < 285$ MeV+ $\mathcal{O}(a)$+ $\mathcal{O}(e^{-mL})$ & 39.38(3.35) & 39.43(3.30) & -0.62(22.83) & 16.7(15)  & 0 \\
no cut in $M_\pi$+ $\mathcal{O}(a)$+ $\mathcal{O}(e^{-mL})$ & 39.34(2.08) &
37.61(2.04) & 22.39(13.53) & 31.1(27)  & 19 \\\hline
$M_\pi < 220$ MeV & 46.81(1.14) & 44.88(1.16) & 24.92(5.61) & 6.9(10)  & 27 \\
$M_\pi < 285$ MeV & 43.71(62) & 42.02(63) & 21.87(3.42) & 27.8(18)  & 2 \\
no cut in $M_\pi$ & 41.04(39) & 39.32(39) & 22.23(2.32) & 92.3(30)  & 0 \\
$M_\pi < 220$ MeV+ $\mathcal{O}(a)$ & 51.38(5.87) & 49.17(5.80) & 28.65(16.12) & 6.3(8)  & 5 \\
$M_\pi < 285$ MeV+ $\mathcal{O}(a)$ & 45.77(2.73) & 44.14(2.73) & 21.17(8.71) & 27.2(16)  & 0 \\
no cut in $M_\pi$+ $\mathcal{O}(a)$ & 40.38(1.65) & 39.02(1.64) & 17.62(4.73) & 90.9(28)  & 0 \\
$M_\pi < 220$ MeV+ $\mathcal{O}(e^{-mL})$ & 47.21(1.20) & 45.28(1.22) & 24.95(5.61) & 6.8(9)  & 10 \\
$M_\pi < 285$ MeV+ $\mathcal{O}(e^{-mL})$ & 44.44(76) & 42.75(77) & 21.79(3.42) & 25.9(17)  & 2 \\
no cut in $M_\pi$+ $\mathcal{O}(e^{-mL})$ & 42.79(56) & 41.08(56) & 22.15(2.32) & 73.4(29)  & 0 \\
$M_\pi < 220$ MeV+ $\mathcal{O}(a)$+ $\mathcal{O}(e^{-mL})$ & 52.26(5.93) & 49.09(6.00) & 41.03(32.57) & 6.0(7)  & 2 \\
$M_\pi < 285$ MeV+ $\mathcal{O}(a)$+ $\mathcal{O}(e^{-mL})$ & 47.13(2.90) & 46.07(2.99) & 13.78(19.15) & 24.6(15)  & 1 \\
no cut in $M_\pi$+ $\mathcal{O}(a)$+ $\mathcal{O}(e^{-mL})$ & 43.83(1.87) & 42.81(1.87) & 13.24(10.25) & 71.9(27)  & 0 \\
	\end{tabular}
	\caption{The results for the different fit variations together with the
	assigned weights $w_i$. Results for the window/two-state data are given
	in the upper/lower panel. For convenience the values have been converted to
physical units using $\sqrt{t_0}$ of Eq.~(\ref{eq:t0physApdx}).}
	\label{tab:extrap_res}
\end{table*}
%\newpage

We note that the AIC averaged result is stable with respect to including models where only
terms of second order in the pion- and kaon-mass are used, and a model adding
polynomial fourth order terms in the chiral counting. The former turns out to have less AIC weight
compared to  our (third-order)  estimate, while the latter
needs to be stabilized using priors. In both cases, the changes in the central values are
insignificant compared to our best estimate, and the error changes within a few
percent, depending
on the prior applied for the fourth-order term. Similarily, removing all data
points with a pion mass above 285 MeV from the analysis only has very small effect on the central
value and error. Moreover, we checked that the
AIC average is also stable against variations in the low-energy constants $F_\phi$, $D$
and $F$. To this end we have varied the values of the LECs 
given in Tab.~11 of Ref.~\cite{RQCD:2022xux} within one standard deviation, and added these as additional models in the
averaging.

\end{document}